\documentclass[english,twocolumn,aps,prl,showpacs]{revtex4}
\usepackage[T1]{fontenc}
\usepackage[latin1]{inputenc}
\usepackage{graphicx}
\usepackage{amssymb}

\makeatletter

\makeatletter

\makeatletter

\input{epsf}

\makeatother

\makeatother

\makeatother

\usepackage{babel}
\makeatother

\begin{document}

\title{Comparative study of strong coupling theories of a trapped Fermi
gas at unitarity}

\author{Hui Hu$^{1,2}$, Xia-Ji Liu$^{2}$, and Peter D. Drummond$^{2}$}

\affiliation{$^{1}$\ Department of Physics, Renmin University of China, Beijing
100872, China \\
 $^{2}$\ ARC Centre of Excellence for Quantum-Atom Optics, Department
of Physics, University of Queensland, Brisbane, Queensland 4072, Australia}

\date{\today{}}

\begin{abstract}
We present a systematic comparison of the most recent thermodynamic
measurements of a trapped Fermi gas at unitarity with predictions
from strong coupling theories and quantum Monte Carlo (MC) simulations.
The accuracy of the experimental data, of the order of
a few percent, allows a precise test of different many-body approaches.
We find that a Nozières and Schmitt-Rink treatment of fluctuations
is in excellent agreement with the experimental data and available
MC calculations at unitarity. 
\end{abstract}


\maketitle
The theory of strongly interacting fermions is of great interest.
Interacting fermions are involved in some of the most important unanswered
questions in condensed matter physics, nuclear physics, astrophysics
and cosmology. Though weakly-interacting fermions are well understood,
new approaches are required to treat strong interactions. In these
cases, one encounters a {}``strongly correlated'' picture which
occurs in many fundamental systems ranging from strongly interacting
electrons to quarks.

The main theoretical difficulty lies in the absence of any small coupling
parameter in the strongly interacting regime, which is crucial for
estimating the errors of approximate approaches. Although there are
numerous efforts to develop strong-coupling perturbation theories
of interacting fermions, notably the many-body \textit{T}-matrix fluctuation
theories \cite{nsr,sademelo,gg94,engelbrecht,ohashi,perali,chen,hld,Combescot,gg07},
their accuracy is not well-understood. Quantum Monte Carlo (QMC) simulations
are also less helpful than one would like, due to the sign problem
for fermions \cite{akkineni} or, in the case of lattice calculations
\cite{bulgac06,burovski}, the need for extrapolation to the zero
filling factor limit.

Recent developments in ultracold atomic Fermi gases near a Feshbach
resonance with widely tunable interaction strength, densities, and
temperatures have provided a unique opportunity to \emph{quantitatively}
test different strong-coupling theories \cite{ohara,kinast,partridge,steward,luo}.
In these systems, when tuned to have an infinite $s$-wave scattering
length - the \emph{unitarity} limit - a simple universal thermodynamic
behavior emerges \cite{universal,natphys}. Due to the pioneering
efforts of many experimentalists, the accuracy of thermodynamic measurements
at unitarity has improved significantly. A breakthrough occurred in
early 2007, when both energy and entropy in trapped Fermi gases were
measured without invoking any specific theoretical model \cite{luo}.
This milestone experiment, arguably the most accurate measurement
in cold atoms, has an accuracy at the level of a few percent.

In this Rapid Communication, using experimental data as a benchmark, we present
an \emph{unbiased} test of several strong coupling theories that are
commonly used in the literature, including QMC simulations. From this
comparison, we show that the simplest theory which incorporates pairing
fluctuations appears to be \emph{quantitatively} accurate at unitarity.
This is the \textit{T}-matrix approximation pioneered by Nozières
and Schmitt-Rink (NSR) \cite{nsr} and others \cite{sademelo,ohashi},
as recently applied to trapped gases in the below threshold superfluid
regime \cite{hld}. We find it describes the observed thermodynamics
extremely well at all temperatures at unitarity, except in regions
very near the superfluid transition temperature $T_{c}$. 

The comparisons show that the simple NSR approximation gives excellent
results. This appears to be related to the important symmetry property
of scale-invariance \cite{wernercastin}, which is a necessary feature
of any exact theory at unitarity, and is shared by the NSR approach.
Our comparative results should therefore be useful in developing new
theoretical approaches for strong interacting fermions, and are relevant
to many fields of physics. In particular, our results might shed light
on the applicability of different $T$-matrix approximations to high-$T_{c}$
superconductors and neutron stars, which are of interest to the condensed
matter and astrophysics communities.

The strong coupling theories that we compare include
several \textit{T}-matrix fluctuation and QMC theories. We refer to
Refs. \cite{bulgac06,burovski} for a detailed description of QMC
methods, and briefly review different \textit{T}-matrix theories.
These involve an infinite set of diagrams --- the ladder sum in the
particle-particle channel. It is generally accepted that this ladder
sum is necessary for taking into account strong pair fluctuations
in the strongly interacting regime, since it is the leading class
of all sets of diagrams \cite{flucttheory}.

The diagrammatic \emph{structure} of different \textit{T}-matrix theories
may be clarified \emph{above} $T_{c}$ for a single-channel model
\cite{chen}, where one writes for the \textit{T}-matrix, $t(Q)=U/[1+U\chi\left(Q\right)]$.
Here and throughout, $Q=(\mathbf{q},i\nu_{n})$, $K=(\mathbf{k},i\omega_{m})$,
while $U^{-1}=m/(4\pi\hbar^{2}a)-\sum_{\mathbf{k}}m/\hbar^{2}\mathbf{k}^{2}$
is the bare contact interaction expressed in terms of the \textit{s}-wave
scattering length. We use $\sum_{K}=$ $k_{B}T\sum_{m}\sum_{\mathbf{k}}$,
where $\mathbf{q}$ and $\mathbf{k}$ are wave vectors, while $\nu_{n}$
and $\omega_{m}$ are bosonic and fermionic Matsubara frequencies.

Different T-matrix fluctuation theories differ in their choice of
the particle-particle propagator \cite{flucttheory}, \begin{equation}
\chi\left(Q\right)=\sum\nolimits _{K}G_{\alpha}(K)G_{\beta}(Q-K),\end{equation}
 and the associated self-energy, $\Sigma(K)=\sum\nolimits _{Q}t(Q)G_{\gamma}(Q-K)$.
The subscripts $\alpha$, $\beta$, and $\gamma$ in the above equations
can either be set to {}``0'', indicating a non-interacting Green's
function $G_{0}(K)=1/[i\omega_{m}-\hbar^{2}\mathbf{k}^{2}/2m+\mu]$,
or be absent, indicating a fully dressed interacting Green's function.
In these cases a Dyson equation, $G(K)=G_{0}(K)/[1-G_{0}(K)\Sigma(K)]$,
is required to self-consistently determine $G$ and $\Sigma$. The
only free parameter, the chemical potential $\mu$, is fixed by the
number equation, $N=2\sum\nolimits _{K}G(K)$.

By taking different combinations of $\alpha$, $\beta$ and $\gamma$,
there are six distinct choices of the \textit{T}-matrix approximation,
for which a nomenclature of $(G_{\alpha}G_{\beta})G_{\gamma}$ will
be used. As noted earlier, there is no known \emph{a priori} theoretical
justification for which is the most appropriate. While having the
same diagrammatic structure, the \textit{T}-matrix approximations
we use above and below $T_{c}$ are computationally different, owing
to the use of different $G_{0}$ (or $G$). Below $T_{c}$, these
Green's functions are 2$\times$2 matrices.

The simplest choice, $(G_{0}G_{0})G_{0}$, was pioneered by NSR above
$T_{c}$ using the thermodynamic potential \cite{nsr}, with a truncated
Dyson equation for $G$, \textit{i.e.}, $G=G_{0}+G_{0}\Sigma G_{0}$.
This theory was extended to the broken-symmetry superfluid phase by
several authors \cite{engelbrecht,ohashi,hld,footnote}, using the
mean-field 2$\times$2 matrix BCS Green's function as {}``$G_{0}$''.
In Ref. \cite{hld}, it was shown that the resulting ground state
energy is in excellent agreement with the zero-temperature QMC calculation
for all interaction strengths. The NSR approximation is the simplest
scheme that includes the effects of particle-particle fluctuations.
It does not attempt to be self-consistent. In the other extreme, one
may consider a $(GG)G$ approximation, with a fully self-consistent
propagator. This was investigated in detail by Haussmann \textit{et
al.}, both above \cite{gg94} and below $T_{c}$ \cite{gg07}. Below
$T_{c}$ an ad-hoc renormalization of the interaction strength is
required to obtain a gapless phonon spectrum.

We also consider an intermediate scheme having an asymmetric form
for the particle-particle propagator, \textit{i.e.}, $(GG_{0})G_{0}$.
This has been discussed in a series of papers by Levin and co-workers
\cite{chen}, based on the assumption that this treatment of fluctuations
is consistent with the simpler BCS theory at low temperatures. Although
the theory has been explored numerically to some extent \cite{levinfull},
a complete numerical solution has not been implemented previously.
A simplified version \cite{chen} of the $(GG_{0})G_{0}$ fluctuation
theory was introduced based on a decomposition of the \textit{T}-matrix
$t(Q)$ in terms of a condensate part and a pseudogap part. In this
Letter, we refer to this approach as the {}``pseudogap model'' and
will include it in our comparative study.

%
\begin{figure}
\begin{centering}
\includegraphics[clip,width=0.4\textwidth]{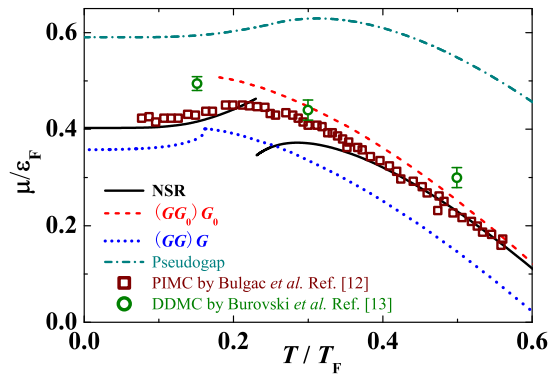} 
\par\end{centering}

\caption{(color online) Chemical potential of a uniform Fermi gas at unitarity
as a function of reduced temperature $T/T_{F}$, where $T_{F}=\epsilon_{F}/k_{B}$.
The lines plotted are the results of NSR (solid lines), $(GG_{0})G_{0}$
(dashed line), $(GG)G$ (dotted line), and pesudogap model (dot-dashed
line). These predictions are compared with lattice QMC simulations
(symbols).}

\label{fig1} 
\end{figure}


Other strong coupling theories with an \emph{artificial} small parameter
have been developed recently, including an $\epsilon$ expansion around
the critical dimension \cite{nishida} and a $1/N$ expansion for
a $2N$-component gas \cite{largeN}. These field-theoretic approaches
provide very useful but so far only qualitative information about
universal thermodynamics valid at unitarity. In the Boltzmann regime
at high temperatures, not explored experimentally so far, it is possible
to make a \emph{virial} expansion in terms of fugacity \cite{hovirial}.
We have verified that the three \textit{T}-matrix schemes we study
here do correctly include the dominant second-order virial contribution
in the high temperature region.

Fig. \ref{fig1} compares the temperature dependence of the chemical
potential at unitarity, calculated from different \textit{T}-matrix
schemes and lattice QMC simulations. The \textit{T}-matrix approximations
above $T_{c}$ have been solved using an adaptive step Fourier transform
method. Below $T_{c}$, the NSR and $(GG)G$ results are from Refs.
\cite{hld} and \cite{gg07}, respectively. The $(GG_{0})G_{0}$
approximation below $T_{c}$ has not been worked out yet. The QMC
results are taken from Refs. \cite{bulgac06} and \cite{burovski}.
However, these lattice calculations may have systematic errors due
to an extrapolation to the zero filling factor limit which is necessary
to have a well-defined continuum theory. Nonetheless, the three \textit{T}-matrix
calculations agree well with the lattice QMC simulations. On the other
hand, the prediction of the pesudogap model above $T_{c}$ differs
substantially from the $(GG_{0})G_{0}$ results, for which it is an
approximation. The pseudogap model omits important features of the
full $(GG_{0})G_{0}$ theory, due to the {}``condensate''+{}``pseudogap''
decomposition of the \textit{T}-matrix.

The determination of energy and entropy is a subtle problem. It is
known that universal thermodynamics at unitarity implies a rigorous
scaling relation \cite{universal}, $P=-\Omega=2E/3$, which relates
the pressure (or thermodynamic potential) and the energy density for
a unitarity gas in the same way as for an ideal, non-interacting quantum
gas, although the energy densities are quite different. Apart from
the $(GG)G$ scheme above-$T_{c}$ and the NSR approach (in both regimes),
strong coupling theories in general \emph{do not} satisfy this essential
scaling relation. The violation is typically at the level of a few
percent, comparable to the accuracy of the experimental data we used.
For quantitative purposes, we calculate the thermodynamic potential
from the chemical potential, using \begin{equation}
\Omega\left(\mu,T=const\right)=-\int_{\mu_{0}}^{\mu}n\left(\mu^{\prime}\right)d\mu^{\prime}+\Omega\left(\mu_{0},T\right)\end{equation}
 at a given temperature. Here, the lower bound of the integral $\mu_{0}$
is sufficiently small so that $\Omega\left(\mu_{0},T\right)$ can
be obtained accurately from a high temperature virial expansion \cite{hovirial}.
The energy and entropy can then be calculated from the rigorous scaling
relations, $E=-3\Omega/2$, and $S=(-5\Omega/2-\mu N)/T$, valid at
unitarity.

%
\begin{figure}
\begin{centering}
\includegraphics[clip,width=0.4\textwidth]{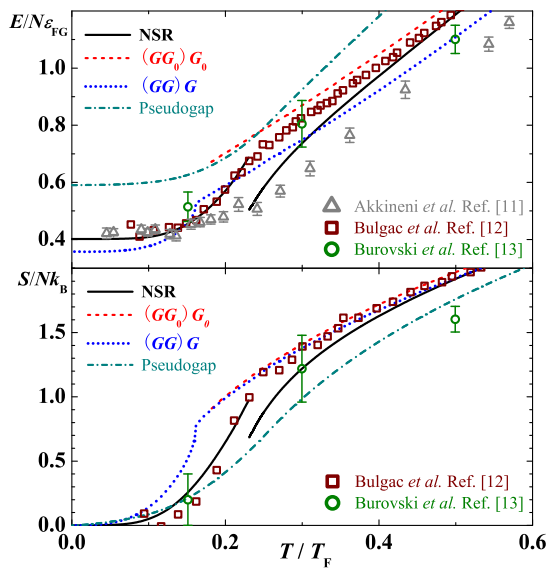} 
\par\end{centering}

\caption{(color online) Temperature dependence of the energy (upper panel)
and of the entropy (lower panel) of a uniform Fermi gas at unitarity,
obtained from different \textit{T}-matrix approximations and QMC simulations
as indicated.}

\label{fig2} 
\end{figure}


The energy and entropy obtained in this manner are given in Fig. \ref{fig2},
and compared to the predictions of QMC calculations. There is a reasonable
agreement between \textit{T}-matrix theories and the lattice QMC simulations.
For the energy, we also show the path-integral Monte Carlo results
of Akkineni \textit{et al.} for the \emph{continuum} model \cite{akkineni}.
At temperatures above $0.2T_{F}$, the energy lies systematically
below that of all the \textit{T}-matrix theories. This is probably
due to the use of a finite effective range $r_{0}$ for the interaction
\cite{akkineni}, \textit{i.e.}, $k_{F}r_{0}\simeq0.3$. Compared
to the QMC results, the pseudogap model appears to provide the least
accurate predictions for energy and entropy. At low temperatures it
predicts a $T^{3/2}$ dependence for the entropy, which is characteristic
of a non-interacting ideal Bose condensed gas. In contrast, the \textit{T}-matrix
entropies follow a $T^{3}$ scaling law, arising from the
Bogoliubov-Anderson phonon modes \cite{gg07}.

We now compare theoretical predictions with experimental data \cite{luo}.
A strongly interacting Fermi gas of $N=1.3(2)\times10^{5}$ lithium
atoms is prepared in a Gaussian trap $V(\mathbf{r})=V_{0}\{1-\exp[-m(\omega_{\perp}^{2}\rho^{2}+\omega_{z}^{2}z^{2})/(2V_{0})]\}$
with $V_{0}\simeq10E_{F}$ at a magnetic field $B=840$ G, slightly
above the resonance position $B_{0}=834$ G. The large but finite
interaction, $k_{F}a=-20.0$, leads to an approximately 1\% correction
that is not accounted for experimentally. The energy is determined
in a model-independent way from the mean square radius $\left\langle z^{2}\right\rangle $
of the strongly interacting cloud, according to the virial theorem,
which states that the potential energy ($\propto\left\langle z^{2}\right\rangle $)
of the gas is a half of its total energy. The entropy of the gas is
calibrated again from the cloud size, but after an adiabatic sweep
to a weakly interacting gas with $k_{F}a=-0.75$, using a precise
theory at weak coupling. We refer to Refs. \cite{luo} and \cite{natphys}
for further details. To determine the energy and entropy theoretically,
we apply the local density approximation (LDA) by assuming that the
system can be treated as locally uniform, with a position-dependent
local chemical potential $\mu_{\hom}[n\left(\mathbf{r}\right),T/T_{F}(n)]=\mu-V(\mathbf{r})$,
where $T_{F}(n)$ is the local Fermi temperature. From this condition,
the density profile is obtained, and the total energy and entropy
are calculated.

%
\begin{figure}
\begin{centering}
\includegraphics[clip,width=0.4\textwidth]{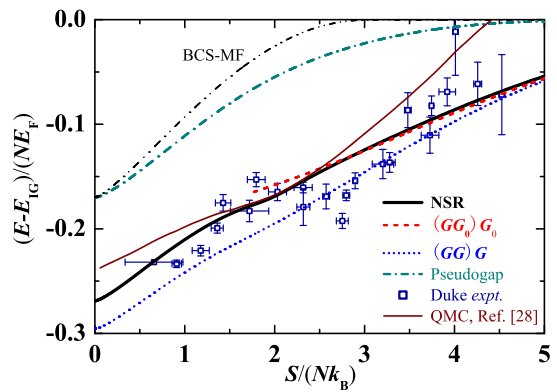} 
\par\end{centering}

\caption{(color online) Theoretically predicted universal thermodynamics in
comparison with experimental data \cite{luo}.}

\label{fig3} 
\end{figure}


Fig. \ref{fig3} shows the interaction energy vs. entropy in a harmonic
trap as predicted by the strong-coupling theories in comparison with
experimental data. All results of perturbation theories, except that
of the NSR approach, were not reported previously to our knowledge.
The energy is expressed in units of the Fermi energy at zero temperature
in a harmonic trap: $E_{F}=(3N\omega_{\perp}^{2}\omega_{z})^{1/3}=k_{B}T_{F}$.
To emphasize the effects of interactions, we have subtracted the ideal
gas result $E_{IG}$. No adjustable parameters have been used theoretically
or experimentally. This comparison is therefore an unbiased test of
how well \textit{T}-matrix theories agree with experiment \cite{luo}.

The difference between different \textit{T}-matrix schemes, mostly
of the order $0.05NE_{F}$, is relatively small and thus nearly indistinguishable
in the plot of total energy and entropy. Despite this, the extraordinary
precision of the measurements is able to discriminate between these
theories in the interaction energy, as given in Fig. \ref{fig3}.
The NSR approach is seen to give the best fit to the experimental
data below $T_{c}$ (corresponding to $S_{c}\approx2.3Nk_{B}$) and
above $T=0.5T_{F}$ (corresponding to $S>3.5Nk_{B}$). This indicates
that the simplest non-self-consistent \textit{T}-matrix approximation
captures the essential physics of strong pair fluctuations at both
low (superfluid) and high (normal) temperatures. Around $T_{c}$,
the experimental data shows evidence of what could be a first-order
superfluid transition. However, due to {}``critical slowing-down'',
systematic experimental errors cannot be ruled out in this regime,
if the magnetic field sweep is not quite adiabatic.

In the temperature region just above $T_{c}$, the NSR approach presumably
does not fully capture the full effect of fluctuations, compared to
the self-consistent $(GG)G$ theory above $T_{c}$. At the transition,
from the experimental data one may determine experimentally a critical
entropy $S_{c}/N\simeq2.3k_{B}$ and a critical energy $E_{c}/N\simeq0.86E_{F}$
in a trap. The critical temperature $T_{c}/T_{F}$ in the case of
a trap is difficult to determine, due to the unknown relation $S(T)$.
The theoretical predictions are 0.29 {[}NSR], 0.21 {[}$(GG_{0})G_{0}$
and $(GG)G$], and 0.27 {[}pseudogap model].

In a further comparison, we include in Fig. \ref{fig3} a recent QMC
calculation (thin solid line) of trapped Fermi gases \cite{bulgac07}.
There is a noticeable systematic difference between the QMC and experimental
data at high entropy, but this is due to the improper use of an ideal
gas approximation in the QMC estimates for large $T$. It is clear
that the unitarity gas remains strongly interacting even close to
the degenerate temperature (\textit{i.e.}, $S\simeq5Nk_{B}$). Thus,
a virial expansion of the equation of state up to the second order
should be applied. Among all pair fluctuating theories, Fig. 3 shows
that the pseudogap approximation gives poor agreement with thermodynamic
data, though it is better than BCS mean-field theory -- which completely
ignores the pairing fluctuations. Therefore, the pseudogap model does
not describe the strong fluctuations at unitarity as well as the full
$(GG_{0})G_{0}$ theory.

In conclusion, the accurate thermodynamic measurements
at Duke University have allowed us to perform a test of strong-coupling
\textit{T}-matrix theories at unitarity. The simplest NSR approximation
for the particle-particle \textit{T}-matrix is found to give the best
quantitative description. Further work is needed to understand the
reason for this, but we conjecture that it is related to scale-invariance
symmetry in the unitarity limit. We conclude that near the BCS-BEC
crossover region, the NSR approximation is surprisingly useful.

We thank Allan Griffin and Edward Taylor for discussions and critical
reading of the manuscript. We also acknowledge V. K. Akkineni et.
al. for providing us with their QMC data. This research is supported
by the Australian Research Council Center of Excellence program, NSF-China
Grant No. 10774190, and NFRP-China Grant Nos. 2006CB921404 and 2006CB921306.


\begin{thebibliography}{10}
\bibitem{nsr}P. Nozières and S. Schmitt-Rink, J. Low Temp. Phys.
\textbf{59}, 195 (1985).

\bibitem{sademelo}C. A. R. Sade Melo, M. Randeria, and J. R. Engelbrecht,
Phys. Rev. Lett. \textbf{71}, 3202 (1993).

\bibitem{gg94}R. Haussmann, Phys. Rev. B \textbf{49}, 12975 (1994).

\bibitem{engelbrecht}J. R. Engelbrecht, M. Randeria, and C. A. R.
Sade Melo, Phys. Rev. B \textbf{55}, 15153 (1997).

\bibitem{ohashi}Y. Ohashi and A. Griffin, Phys. Rev. Lett. \textbf{89},
130402 (2002); Phys. Rev. A \textbf{67}, 063612 (2003).

\bibitem{perali}A. Perali \textit{et al.}, Phys. Rev. Lett. \textbf{92},
220404 (2004).

\bibitem{chen}B. R. Patton, Ph.D. thesis, Cornell University, 1971 (unpublished);
Q. J. Chen \textit{et al.}, Phys. Rep. \textbf{412}, 1 (2005).

\bibitem{hld}H. Hu, X.-J. Liu, and P. D. Drummond, Europhys. Lett.
\textbf{74}, 574 (2006); Phys. Rev. A \textbf{73}, 023617 (2006);
R. B. Diener, R. Sensarma, and M. Randeria, Phys. Rev. B \textbf{77},
023626 (2008).

\bibitem{Combescot}R. Combescot, X. Leyronas and M.Yu. Kagan, Phys.
Rev. A \textbf{73}, 023618 (2006); Z. Nussinov and S. Nussinov, Phys.
Rev. A \textbf{74}, 053622 (2006) .

\bibitem{gg07}R. Haussmann \textit{et al.}, Phys. Rev. A \textbf{75},
023610 (2007).

\bibitem{akkineni}V. K. Akkineni, D. M. Ceperley, and N. Trivedi,
Phys. Rev. B \textbf{76}, 165116 (2007).

\bibitem{bulgac06}A. Bulgac, J. E. Drut, and P. Magierski, Phys.
Rev. Lett. \textbf{96}, 090404 (2006).

\bibitem{burovski}E. Burovski \textit{et al.}, Phys. Rev. Lett. \textbf{96},
160402 (2006).

\bibitem{ohara}K. M. O'Hara \textit{et al.}, Science \textbf{298},
2179 (2002).

\bibitem{kinast}J. Kinast \textit{et al.}, Science \textbf{307},
1296 (2005).

\bibitem{partridge}G. B. Partridge \textit{et al.}, Science \textbf{311},
503 (2006).

\bibitem{steward}J. T. Steward \textit{et al.}, Phys. Rev. Lett.
\textbf{97}, 220406 (2006).

\bibitem{luo}L. Luo \textit{et al}., Phys. Rev. Lett. \textbf{98},
080402 (2007).

\bibitem{universal}T.-L. Ho, Phys. Rev. Lett. \textbf{92}, 090402
(2004); J. E. Thomas, J. Kinast, and A. Turlapov, \textit{ibid}. \textbf{95},
120402 (2005).

\bibitem{natphys}H. Hu, P. D. Drummond, and X.-J. Liu, Nat. Phys.
\textbf{3}, 469 (2007).

\bibitem{wernercastin}F. Werner and Y. Castin, Phys. Rev. A \textbf{74},
053604 (2006).

\bibitem{flucttheory}V. M. Loketev, R. M. Quick, and S. G. Sharapov,
Phys. Rep. \textbf{349}, 1 (2001).

\bibitem{footnote}Different realization of the NSR approaches in
the superfluid phase below $T_{c}$ are reviewed by Taylor and Griffin;
see, E. Taylor, PhD thesis, University of Toronto (2007).

\bibitem{levinfull} J. Maly, B. Janko, and K. Levin, Physica C \textbf{321},
113 (1999); Phys. Rev. B \textbf{59}, 1354 (1999).

\bibitem{nishida}Y. Nishida, Phys. Rev. A \textbf{75}, 063618 (2007).

\bibitem{largeN}P. Nikoli\'{c} and S. Sachdev, Phys. Rev. A \textbf{75},
033608 (2007); M. Y. Veillette, D. E. Sheehy, and L. Radzihovsky,
\textit{ibid}. \textbf{75}, 043614 (2007).

\bibitem{hovirial}T.-L. Ho and E. J. Mueller, Phys. Rev. Lett. \textbf{92},
160404 (2004).

\bibitem{bulgac07}A. Bulgac, J. E. Drut, and P. Magierski, Phys.
Rev. Lett. \textbf{99}, 120401 (2007). 
\end{thebibliography}
\end{document}